\documentclass[11pt,a4paper,citesort]{article}
\usepackage[makeroom]{cancel}
\usepackage{bm}
\usepackage{amsmath}
\usepackage{amsfonts}
\usepackage{graphicx}
\usepackage{nicefrac}
\usepackage{authblk}
\usepackage{cite}
\usepackage[left=2.54cm,top=2.54cm,right=2.54cm,bottom=2.54cm,nohead]{geometry}
\DeclareGraphicsExtensions{.png,.jpg,.pdf}
\usepackage{epstopdf}
\usepackage{float}
\usepackage{subfigure}
\usepackage{fouriernc}

\title{Analysis of electrolyte transport through charged nanopores}

\usepackage{authblk}
\makeatletter
\renewcommand\AB@authnote[1]{\textsuperscript{\normalfont#1}}

\makeatother

\author[1,2]{P.B. Peters,}
\author[1]{R. van Roij,}
\author[3,4]{M.Z. Bazant}
\author[5,6]{P.M. Biesheuvel}
\affil[1]{Institute for Theoretical Physics, Center for Extreme Matter and Emergent Phenomena, Utrecht University, Leuvenlaan 4, 3584 CE Utrecht, The Netherlands.}
\affil[2]{Fitzwilliam College, University of Cambridge, Cambridge CB3 0DG, United Kingdom.}
\affil[3]{Departments of Chemical Engineering and Mathematics, Massachusetts Institute of Technology, Cambridge, MA 02139, USA.}
\affil[4]{Department of Materials Science and Engineering and SUNCAT Center of Interfacial Science and Catalysis, Stanford University, Stanford, CA 94305, USA.}
\affil[5]{Wetsus, European Centre of Excellence for Sustainable Water Technology, Oostergoweg 9, 8911 MA Leeuwarden, The Netherlands.}
\affil[6]{Laboratory of Physical Chemistry and Soft Matter, Wageningen University, Dreijenplein 6, 6703 HB Wageningen, The Netherlands.}

\date{} 

\begin{document}

\maketitle

\begin{abstract}
We revisit the classical problem of flow of electrolyte solutions through charged capillary nanopores or nanotubes as described by the capillary pore model (also called ``space charge'' theory). This theory assumes very long and thin pores and uses a one-dimensional flux-force formalism which relates fluxes (electrical current, salt flux, fluid velocity) and driving forces (difference in electric potential, salt concentration, pressure).  
We analyze the general case with overlapping electric double layers in the pore and a nonzero axial salt concentration gradient. The 3 $\times$ 3 matrix relating these quantities exhibits Onsager symmetry and we report a significant new simplification for the diagonal element relating axial salt flux to the gradient in chemical potential. We prove that Onsager symmetry is preserved under changes of variables, which we illustrate by transformation to a different flux-force matrix given by Gross and Osterle (1968). The capillary pore model is well-suited to describe the nonlinear response of charged membranes or nanofluidic devices for electrokinetic energy conversion and water desalination, as long as the transverse ion profiles remain in local quasi-equilibrium.  As an example, we evaluate electrical power production from a salt concentration difference by reverse electrodialysis, using an efficiency vs. power diagram. We show that since the capillary pore model allows for axial gradients in salt concentration, partial loops in current, salt flux or fluid flow can develop in the pore. Predictions for macroscopic transport properties using a reduced model where the potential and concentration are assumed to be invariant with radial coordinate (``uniform potential'' or ``fine capillary pore'' model), are close to results of the full model. 
\end{abstract}

\section{Introduction}

\begin{figure}[H]
\centering
\includegraphics[scale=0.55]{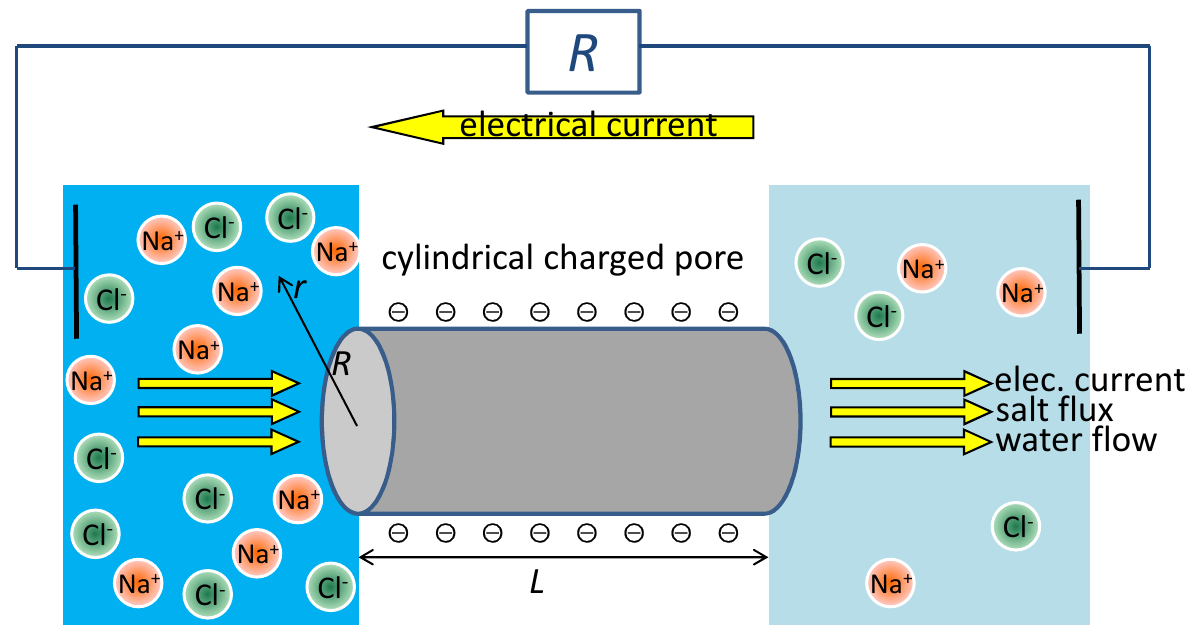}
\caption{Sketch of a charged cylindrical pore subjected to water flow, electrical current and salt flux between a high salinity (left) and low-salinity (right) reservoir.}
\label{fig:Poremodel}
\end{figure}

Charged capillary nanopores and nanotubes are essential in many natural and technological systems, as part of porous membranes separating two aqueous electrolytes ~\cite{Teorell,Gross&Osterle,Fair&Osterle, Sasidhar&Ruckenstein,Sasidhar&Ruckenstein_II,Neogi&Ruckenstein, Wang,Heyden,Hatlo,YanJPCC,Kang,yaroshchuk2012cp,zholkovskij2015}. Membranes containing charged nanopores can be used for water desalination, selective ion removal, and electrokinetic energy conversion. In steady-state, transport is defined by three fluxes (salt flux, electrical current, fluid velocity) and three driving forces (salt concentration difference, electric potential difference, pressure difference). In any physical situation, three out of these six fluxes or forces are required (prescribed) to fully define the problem, with the other three physical quantities to be measured or calculated. It is also possible that one of the three defining relations includes a combination of factors, such as a relation between current and electric potential difference (i.e., applying a constant external electrical load). 
The theory for charged capillaries dates back to the work of Osterle and coworkers \cite{Gross&Osterle, Fair&Osterle} and describes the flow of ions and water through a cylindrical pore carrying a homogeneous charge on its inner surface. The pore is connected to two reservoirs having different salt concentration, pressure and/or electric potential. Length $L$ of the pore is assumed to be many time larger than pore radius $R$. The physical situation is illustrated in Fig. \ref{fig:Poremodel}.   

Although this problem was first analyzed as a simple model for electrokinetic phenomena in membranes, recent interest has also been driven by other  applications, such as electro-osmotic micropumps~\cite{yao2003a,yao2003b,yang2003} and nanofluidic devices~\cite{bruus_book,schoch2008}. In the latter case, the ideal geometry of a straight nanochannel is easily realized in experiments, albeit usually with a rectangular cross section.  Applications of electrokinetic phenomena in nanochannels include streaming current measurements~\cite{werner2001,stein2004,heyden2005}, electrokinetic energy conversion~\cite{morrison1965,Heyden,heyden2007}, ionic~\cite{plecis2005,karnik2005} and flow~\cite{ghowsi1991,schasfoort1999} field-effect transistors,  electro-osmotic impedance effects~\cite{schiffbauer2013}, and electrophoretic separations~\cite{pennathur2005a,pennathur2005b,yaroshchuk2011,yaroshchuk2012}.    

Until recently, most of the theoretical literature on membranes and nanochannels has been based on the assumptions of thin  electric double layers (EDLs), negligible axial salt concentration gradients, and local quasi-equilibrium of the ion distributions in the potential. It is well known that interfaces between charged membranes or nanochannels and  unsupported bulk electrolytes lead to ion concentration polarization outside the membrane, e.g., in classical electrodialysis~\cite{Sonin,nikonenko2010,Tedesco}, but complex non-equilibrium electrokinetic phenomena resulting from strong concentration polarization have recently been discovered {\it inside} membrane pores or microchannels, such as deionization shock waves~\cite{mani2009,zangle2009,zangle2010,mani2011,yaroshchuk2012,dydek2013,nam2015} and over-limiting current sustained by surface conduction (electro-migration) and electro-osmotic flow~\cite{dydek2011,yaroshchuk2011EOF,nielsen2014,deng2013} with applications to nano-templated electrodeposition~\cite{han2014} and water desalination by ``shock electrodialysis''~\cite{schlumpberger2015}.   In most situations for nanochannels, the ions remain in local quasi-equilibrium, since electro-migration and diffusion dominate, although non-equilibrium structures, such as ``salt fingers'' extending along the pore surfaces, can arise in microchannels, if electro-osmotic convection dominates~\cite{dydek2011,rubinstein2013,nielsen2014,nam2015}.  Here, we neglect such effects and focus on deriving the consequences of local (but not global) quasi-equilibrium.

In this work we revisit Osterle's capillary pore model  \cite{Gross&Osterle, Fair&Osterle} describing the nonlinear  electrokinetic response of charged nanopores (also used by Sasidhar and Ruckenstein \cite{Sasidhar&Ruckenstein}). We show that all three-fold integrals in the theory (which must be evaluated across the pore radius) can be simplified to single integrals, thereby significantly simplifying numerical calculations. We also demonstrate how an infinite number of local flux-force relationships are in principle possible that all abide Onsager symmetry. Unlike most prior work, we consider the general case where EDLs overlap, and axial gradients in salt concentration are not negligible. Note that even when the external bulk solutions have equal salt concentration, concentration polarization due to current or flow leads to a concentration difference between the pore ends~\cite{YanJPCC}. The full capillary pore model therefore allows us to describe reverse electrodialysis, a membrane process to extract electrical energy from salinity differences, e.g., between river water and seawater~\cite{Fair&Osterle,post2008,dlugolecki2009,Siria,Lee_energies,Rankin}.  We provide numerical results for energy conversion, and of two-dimensional (axisymmetric) current profiles for pores with EDL overlap in the presence of an overall salt concentration difference. Though we present only calculation results for the steady-state, the model can be extended quite straightforwardly to dynamic situations \cite{Keh&Tseng}. In the model, ions are assumed to be fully dissociated monovalent point charges. Theory for electrolytes containing ampholytic ions (ions that can undergo acid/base reactions) is discussed in refs.~\cite{Jacobs&Probstein, Sounart&Baygents, Dykstra}. 

\section{General theory}

\subsection{Governing equations}
The derivation that follows closely resembles that of Gross and Osterle \cite{Gross&Osterle}, Fair and Osterle \cite{Fair&Osterle} and Sasidhar and Ruckenstein \cite{Sasidhar&Ruckenstein}. Central to the theory are three equations: the extended Navier-Stokes (NS) equation, the extended Nernst-Planck (NP) equation and the Poisson equation. The NP-equation describes the molar flux $\bm{J}_i$ (mol/m$^2$/s) of ions of species $i$ by 

\begin{equation}\label{eq:NernstPlanck}
\bm{J}_i(x,r) = c_i(x,r) \bm{u}(x,r) - D_i \left(\nabla c_i(x,r)+\frac{z_i c_i}{\Phi_\text{B}}\nabla\Phi(x,r)\right)
\end{equation}
where $c_i(x,r)$ is local ion concentration (mM=mol/$\text{m}^{3}$), $\Phi (x,r)$ local electric potential (V), $\Phi_\text{B}$ thermal voltage $(=R_g T/F)$, and $D_i$ the diffusion coefficient of species i ($\text{m}^2/\text{s}$) with $i\in$\{+,-\}. Ion valency $z_i$ is either $+1$ or $-1$ because we will consider only a 1:1 salt (with ions, e.g., Na$^+$ and Cl$^-$). Further, $\bm{u}$ is velocity of the fluid (m/s) and $T$ temperature (K). Faraday's constant is $F =$ 96485 C/mol and the gas constant is $R_g=8.3144$ J/mol/K. Eq. (\ref{eq:NernstPlanck}) assumes that ions are volumeless point charges. 
In this work we consider a stationary state and thus the ion mass balance 
\begin{equation}\label{eq:continuity}
\frac{\partial c_i(x,r)}{\partial t} + \nabla\cdot \bm{J}_i(x,r) = 0
\end{equation}
simplifies to $\nabla\cdot \bm{J}_i(x,r) = 0$. 
Throughout we assume cylindrical symmetry, see Fig. \ref{fig:Poremodel}, with axial coordinate $x\in[0,L]$ and radial coordinate $r\in[0,R]$. 
 
For laminar flow, fluid flow is described by the incompressible NS-equation, which at the low Reynolds number of interest here is given by
\begin{equation}\label{eq:NavierStokes}
\mu\nabla^2\bm{u}(x,r) - \nabla p_h(x,r) - \rho(x,r)\nabla\Phi(x,r) = 0 \quad\textrm{and}\quad \nabla\cdot\bm{u}(x,r)=0
\end{equation}
where $\mu$ is viscosity (Pa$\cdot$s), $p_h(x,r)$ hydrostatic pressure (Pa), and $\rho(x,r)$ local charge density (C/m$^3$).

Finally, Poisson's equation relates potential $\Phi(x,r)$ to charge density as
\begin{equation}\label{eq:Poisson}
\nabla^2\Phi(x,r) = -\frac{\rho(x,r)}{\varepsilon} = -\frac{F}{\varepsilon}(c_+(x,r) - c_-(x,r))
\end{equation}
where $\varepsilon$ is the permittivity of the medium (F/m). In the second equality of Eq. (\ref{eq:Poisson}) we implement the assumption of a 1:1 salt (both ions monovalent), and only consider positions $0<r<R$ away from the surface charge of the pore at $r=R$.

\subsection{Boundary conditions \& further assumptions}

Because the pore is much longer than wide, we can assume local equilibrium in the $r$-direction and decompose the total potential as~\cite{Koh&Anderson}
\begin{equation}\label{eq:decomposition}
\Phi(x,r) = \phi_v(x) + \psi(x,r)
\end{equation}
where the ``radial potential'' $\psi(x,r)$ is obtained from an equilibrium PB-model,
and $\phi_v(x)$ accounts for axial gradients in potential (along the length of the pore).
Concerning the fluxes $\bm{J}_i$ and $\bm{u}$, the walls of the pore are impermeable to both fluid and ions, so we have
\begin{equation}\label{eq:impermeable}
J_{i,r}(x,R)= 0 \hspace{1.2em} ,\quad u_r(x,R) = 0
\end{equation}
where subscript $r$ denotes the radial component of vector quantities $\bm{J}_i(x,r)$ and $\bm{u}(x,r)$.
We also assume no-slip boundary conditions for fluid velocity $\bm{u}$, i.e.,
\begin{equation}\label{no-slip}
u_x(x,R)=0
\end{equation}
where we stress that this does not hold for ion fluxes $\bm{J}_i$. 

Naturally, the system we have described is out of equilibrium and to account for this, ``virtual'' quantities are defined, which express the principle of local equilibrium~\cite{baldessari2008,yaroshchuk2011}. A physical quantity $F_v(x)$ (subscript ``$v$'') is defined as ``virtual'' when it represents conditions in a virtual reservoir that is in equilibrium with any differential volume (``slice'') in the pore. Thus, it represents conditions under which a cylindrical pore cross-section, or slice, is in equilibrium with a charge neutral reference volume. From this definition it follows that virtual quantities will be $x$-dependent only. Virtual properties at the two ends of the pore correspond to conditions just outside the pore in bulk solution \cite{Westermann-Clark&Anderson}. In the capillary pore model we encounter virtual concentration $c_v(x)$, virtual pressure $p_{t,v}(x)$ and virtual potential $\phi_v(x)$.
With this formalism defined, we can impose the most important assumption, namely that the pore is much longer than wide, or $L\gg R$, with $L$ pore length and $R$ pore radius. We are then allowed to assume the ionic profile inside the pore to be in local equilibrium in radial direction \cite{Grossthesis} leading to 
\begin{equation}\label{eq:r-equilibrium}
J_{i,r}(x,r)= 0 \hspace{1.2em} ,\quad u_r(x,r) = 0
\end{equation} 
allowing us to derive a radial PB-equation by inserting Eqs. (\ref{eq:decomposition}) and (\ref{eq:r-equilibrium}) in the $r$-component of Eq. (\ref{eq:NernstPlanck}) which results in 
\begin{equation}
\frac{\partial c_i(x,r)}{\partial r} = -\frac{z_i c_i(x,r)}{\Phi_\text{B}} \frac{\partial\psi(x,r)}{\partial r}
\end{equation}
which can be integrated to the Boltzmann distribution
\begin{equation}\label{eq:Boltzmann}
c_i(x,r) = c_v(x) \>\exp\left(-z_i \frac{\psi(x,r)}{\Phi_\text{B}}\right)
\end{equation}
and implemented in Eq. (\ref{eq:Poisson}) to obtain the desired PB-equation
\begin{equation}\label{eq:PoissonBoltzmann}
\nabla^2\Phi(x,r) = 2 \frac{F c_v(x)}{\varepsilon} \sinh\left(\frac{\psi(x,r)}{\Phi_\text{B}}\right)
\end{equation}
which can be solved with boundary conditions of fixed charge and considering cylindrical symmetry,
\begin{equation}\label{eq:BoundaryPB}
\left.\frac{\partial\psi(x,r)}{\partial r}\right|_{r=R} = {}+{}\frac{\sigma}{\varepsilon} \hspace{1.2em} , \quad \left.\frac{\partial\psi(x,r)}{\partial r}\right|_{r=0} = 0
\end{equation}
where $\sigma$ is surface charge density of the pore wall (in C/m$^2$). In the present work, $\sigma$ is assumed constant, invariant along the pore. However, in reality it will often depend on the local pH in the pore via a surface ionization mechanism and thus gradients in $\sigma$ can develop~\cite{Catalano_Arxiv}. 

\subsection{Non-dimensional formulation}

In order to simplify our governing equations (\ref{eq:NernstPlanck}), (\ref{eq:NavierStokes}) and (\ref{eq:PoissonBoltzmann}) it is convenient to non-dimensionalize all physical quantities by division with an appropriate reference quantity. This change of variables is listed below and with a slight abuse of notation we replace all variables by their dimensionless counterparts, as follows,
\begin{align} \label{eq:ChangeofVariables}
\frac{r}{R} &\rightarrow r & \frac{x}{L} &\rightarrow x   \nonumber\\
\frac{\phi_v(x)}{\Phi_\text{B}}&\rightarrow \phi_v(x) & \frac{\psi(x,r)}{\Phi_\text{B}} &\rightarrow \psi(x,r) \nonumber\\ 
\frac{p_h(x,r)}{p_\mathrm{ref}}&\rightarrow p_h(x,r)  & \frac{c_v(x)}{c_\mathrm{ref}} &\rightarrow c_v(x) & &   \nonumber\\
\frac{\bm{J}_i(x,r)}{J_\mathrm{ref}} &\rightarrow \bm{j}_i(x,r) & \frac{\bm{u}(x,r)}{u_\mathrm{ref}} &\rightarrow \bm{u}(x,r)\\
\frac{\sigma}{\sigma_\mathrm{ref}} &\rightarrow \sigma & u_\mathrm{ref}&= \frac{D}{L}  \nonumber\\
\Phi_\text{B} &= \frac{R_g T}{F} & p_\mathrm{ref}&= c_\mathrm{ref} R_g  T  \nonumber\\
J_\mathrm{ref}&=\frac{D c_\mathrm{ref}}{L} &\sigma_\mathrm{ref} &= \frac{\varepsilon \Phi_\text{B}}{R} \nonumber
\end{align}
where $c_\mathrm{ref}$ is an arbitrary reference concentration, for which we use $c_\mathrm{ref}=1$ mM (1 mol/m$^3$) and where $D$ is the (assumed equal) diffusion coefficient of both types of ions. With this change of variables we now have $r\in[0,1]$ and $x\in[0,1]$. Like hydrostatic pressure, $p_h$, other (virtual) pressures to be introduced below are also scaled to reference pressure $p_\mathrm{ref}$. From this point onward, all equations and parameters are non-dimensional, unless otherwise stated.

Next we proceed with the approximation that in the limit $L\gg R$ we can ignore the $\frac{\partial^2}{\partial x^2}$ terms in both the NS-equation (\ref{eq:NavierStokes}) and PB-equation (\ref{eq:PoissonBoltzmann}), which is a well-known procedure \cite{Grossthesis}.
Eq. (\ref{eq:PoissonBoltzmann}) can now be written as
\begin{equation} \label{eq:PoissonBoltzmanndim}
\frac{1}{r}\frac{\partial}{\partial r}\left(r\frac{\partial \psi(x,r)}{\partial r}\right) = \frac{c_v(x)}{\lambda_\mathrm{ref}^2}\sinh\psi(x,r)
\end{equation}
where
\begin{equation}
\lambda_\mathrm{ref}=\frac{1}{R}\textrm{ }\sqrt{\frac{\varepsilon \Phi_\text{B}}{2 F c_\mathrm{ref}}}
\end{equation}
is a dimensionless reference Debye length in units of the cylinder radius, $R$.
Boundary conditions of the PB-equation become \cite{Neogi&Ruckenstein}
\begin{equation}\label{eq:BoundaryPB_dimless}
\left.\frac{\partial\psi(x,r)}{\partial r}\right|_{r=1} = +\sigma \hspace{1.2em} , \quad \left.\frac{\partial\psi(x,r)}{\partial r}\right|_{r=0} = 0
\end{equation}
while the Boltzmann distribution of Eq. (\ref{eq:Boltzmann}) is now written as
\begin{equation}\label{eq:Boltzmanndim}
c_i(x,r)=c_v(x)\> \exp(-z_i \psi(x,r)).
\end{equation}
Performing the change of variables in Eq (\ref{eq:ChangeofVariables}), we can also simplify the NP-equation in the $x$-direction (which is the only direction of interest for the ion fluxes), resulting in 
\begin{equation} \label{eq:NernstPlanckdim}
j_{i,x}(x,r) = c_i(x,r) u_x(x,r) - \frac{\partial c_i(x,r)}{\partial x} - z_i c_i(x,r)  \frac{\partial (\psi(x,r) + \phi_v(x))}{\partial x}
\end{equation}
and simplify the NS-equation by ignoring the $\frac{\partial^2}{\partial x^2}$ terms and substituting Eqs. (\ref{eq:Poisson}), (\ref{eq:ChangeofVariables}), and  (\ref{eq:Boltzmanndim}) in Eq. (\ref{eq:NavierStokes}). In $x$-direction we find that
\begin{equation}\label{eq:NavierStokesdim}
\alpha\frac{1}{r}\frac{\partial}{\partial r}\left(r \frac{\partial u_x(x,r)}{\partial r}\right) - \frac{\partial p_h(x,r)}{\partial x} + 2 c_v(x) \sinh\psi(x,r)\frac{\partial (\psi(x,r) + \phi_v(x))}{\partial x} = 0 
\end{equation}
with the dimensionless viscosity parameter $\alpha$ given by 
\begin{equation}
\alpha= \frac{\mu D}{c_\mathrm{ref}R_g T R^2}.
\end{equation}

\section{Radially averaged flux-force relationships}

In a next step, mathematical expressions are derived for the radially averaged $x$-component of the fluxes, which in case of a flux component $f_x(x,r)$ takes the form
\begin{equation}\label{eq:poreaverage}
 \overline{f_x(x)} = 2\int^1_0 r\cdot f_x(x,r) dr.
\end{equation}
Our aim will be to derive Onsager relations between the radially averaged $x$-component of fluxes $\overline{u_x(x)}$, $\overline{j_{\mathrm{ions},x}(x)}$  and $\overline{j_{\mathrm{ch},x}(x)}$, and driving forces $-\partial_x p_{t,v}(x)$, $-\partial_x \mu_v(x)$ and $-\partial_x\phi_v(x)$.
Here, ion flux $\bm{j}_\mathrm{ions}$ and ionic current $\bm{j}_\mathrm{ch}$ are defined as 
\begin{align} \label{eq:ionflux}
\begin{split}
\bm{j}_\mathrm{ions}(x,r) &=\bm{j}_+(x,r) + \bm{j}_-(x,r)\\
\bm{j}_\mathrm{ch}(x,r) &=\bm{j}_+(x,r) - \bm{j}_-(x,r).
\end{split}
\end{align}
Furthermore, virtual chemical potential, $\mu_v(x)$, virtual osmotic pressure, $\pi_v(x)$, and virtual total pressure, $p_{t,v}(x)$, are defined as
\begin{equation} \label{eq:chemicalpotential}
\begin{alignedat}{2}
\mu_v(x) & = &\enspace\ln c_v&(x)\\
\pi_v(x) & = &2\>c_v&(x)\\
 p_{t,v}(x) & = & p_{h,v}&(x) - \pi_v(x).
\end{alignedat}
\end{equation}
To simplify notation, from this point onward, the $x$-dependency of the various quantities will no longer be explicitly stated.
Inserting the NP-equation (\ref{eq:NernstPlanckdim}) and the Boltzmann distribution (\ref{eq:Boltzmanndim}) into Eq. (\ref{eq:ionflux}), while also using the definitions of Eq. (\ref{eq:chemicalpotential}), we immediately obtain an explicit expression for ion flux and ionic current in $x$-direction, namely
\begin{equation} \label{eq:j_ions}
\begin{alignedat}{4}
j_{\mathrm{ions},x}(r)  &=&2c_v\cosh\psi(r)&u_x(r) &{}-{} 2  c_v \cosh\psi(r)&\frac{\partial \mu_v}{\partial x} &{}+{} 2 c_v \sinh\psi(r)&\frac{\partial\phi_v}{\partial x} \\
j_{\mathrm{ch},x}(r)  &=& {}-{}2c_v \sinh\psi(r)&u_x(r)&  {}+{} 2 c_v \sinh\psi(r)&\frac{\partial \mu_v}{\partial x} &{}-{} 2 c_v \cosh\psi(r)&\frac{\partial\phi_v}{\partial x}.
\end{alignedat}
\end{equation}

We now proceed to find an expression for $u_x(r)$. To this end we note that in $r$-direction the dimensionless NS-equation becomes (using Eq. (\ref{eq:Boltzmanndim}) and $u_r(r)=0$ in Eq. (\ref{eq:r-equilibrium}))
\begin{equation} \label{eq:hpressure}
\frac{\partial p_h(r)}{\partial r} = - \rho(r) \frac{\partial \psi(r)}{\partial r}
=2 c_v\sinh\psi(r) \frac{\partial \psi(r)}{\partial r}
= 2 c_v \frac{\partial \cosh\psi(r)}{\partial r}
\end{equation}
which can again be integrated to result in
\begin{equation} \label{eq:pressure}
p_h(r) - p_{h,v}= 2 c_v \left(\cosh\psi(r) - 1 \right).
\end{equation}
Now we observe that by Eqs. (\ref{eq:hpressure}) and (\ref{eq:pressure}), the equation
\begin{equation}
\frac{\partial p_h(r)}{\partial x}= \frac{\partial p_{t,v}}{\partial x} + 2 c_v\cosh\psi(r) \frac{\partial \mu_v}{\partial x} + 2 c_v\sinh\psi(r)\frac{\partial \psi(r)}{\partial x}
\end{equation}
should hold. Substituting this result back into the NS-equation (\ref{eq:NavierStokesdim}) for the $x$-direction, we arrive at 
\begin{equation}
\alpha\frac{1}{r}\frac{\partial}{\partial r}\left(r \frac{\partial u_x(r)}{\partial r}\right) = \frac{\partial p_{t,v}}{\partial x} + 2 c_v\cosh\psi(r) \frac{\partial \mu_v}{\partial x} -2  c_v \sinh\psi(r) \frac{\partial\phi_v}{\partial x}. 
\end{equation}
Using the fact that $\partial_r u_x(0)=0$, multiplying both sides by $r$ and integrating, we now find 
\begin{align}\label{eq:partial u}
\alpha r \frac{\partial u_x(r)}{\partial r}
=\frac{1}{2} r^2\frac{\partial p_{t,v}}{\partial x} + 2 c_v \int^r_0 r' \cosh\psi(r') dr' \frac{\partial \mu_v}{\partial x} - 2 \lambda_\mathrm{ref}^2 \> r \> \frac{\partial\psi(r)}{\partial r} \frac{\partial\phi_v}{\partial x}
\end{align}
where we reduced the last term by virtue of the identity
\begin{equation}
2 c_v\int^r_0 r' \sinh\psi(r') dr' = 2 \lambda_\mathrm{ref}^2 \int^r_0\frac{\partial}{\partial r'}\left(r'\frac{\partial\psi(r')}{\partial r'}\right) dr' = 2 \lambda_\mathrm{ref}^2 \> r \> \frac{\partial\psi(r)}{\partial r} 
\end{equation}
in which the PB-equation (\ref{eq:PoissonBoltzmanndim}) is implemented. Finally, dividing both sides of Eq.  (\ref{eq:partial u}) by $r$ and using $u_x(1)=0$ we obtain 
\begin{align}\label{eq:u-expression}
\alpha u_x(r)
=-\frac{1}{4} \left(1-r^2 \right) \frac{\partial p_{t,v}}{\partial x} 
- 2 c_v \int^1_r \frac{1}{r_1}\int^{r_1}_0 r_2 \cosh\psi(r_2) dr_2 dr_1  \frac{\partial \mu_v}{\partial x}
-2\lambda_\mathrm{ref}^2 \left(\psi(r) - \psi_\mathrm{w} \right) \frac{\partial\phi_v}{\partial x}
\end{align}
where $\psi_\text{w}$ is the value of potential $\psi$ at the pore wall. It is now a straightforward endeavor to insert $u_x(r)$ back into Eq. (\ref{eq:j_ions}) and take the average defined by Eq. (\ref{eq:poreaverage}). The final result (after grouping all terms) can be written as a matrix equation, relating fluxes, 
\def\yPh{\vphantom{j_{\mathrm{ch},x}}}
$\overline{u_x \yPh}$, $\overline{j_{\mathrm{ions},x}}$  and $\overline{j_{\mathrm{ch},x}}$, and
driving forces, $-\partial_x p_{t,v}$, $-\partial_x \mu_v$ and $-\partial_x\phi_v$, according to 
\begin{equation}\label{eq:fluxesnforces}
\left( \overline{u_x \yPh} \>,\> \overline{j_{\mathrm{ions},x}} \>,\> \overline{j_{\mathrm{ch},x}} \right)^\textrm{t}=\begin{pmatrix} L_{11} & L_{12} & L_{13}\\  L_{21} & L_{22} & L_{23}\\ L_{31} & L_{32} & L_{33} \end{pmatrix}\cdot \left(  -\frac{\partial p_{t,v}}{\partial x}\>,\>-\frac{\partial \mu_v}{\partial x} \>,\>  -\frac{\partial\phi_v}{\partial x} \right)^\textrm{t}
\end{equation}
where the coefficients of this $L$-matrix are either constant or only dependent on the $x$-coordinate, and given by
\begin{equation}\label{eq:Lmatrix}
\begin{alignedat}{2}
L_{11}&=+\,\frac{1}{8\alpha}&&\\
L_{12}&=+\frac{4 c_v}{\alpha} &&\int^1_0 r \int^1_r \frac{1}{r_1}\int^{r_1}_0 r_2 \cosh\psi(r_2) dr_2 dr_1 dr \\
L_{13}&=+\hspace{0.35em}\frac{4}{\alpha} &&\int^1_0 r\lambda_\mathrm{ref}^2 \left(\psi(r) - \psi_\mathrm{w}\right)dr \\
L_{21}&=+\;\frac{c_v}{\alpha} &&\int^1_0 \left(r-r^3 \right)\cosh\psi(r)dr\\
L_{22}&=+\frac{8 c_v}{\alpha} && \int^1_0 r \cosh\psi(r) \left(c_v\int^1_r \frac{1}{r_1}\int^{r_1}_0 r_2 \cosh\psi(r_2) dr_2 dr_1 + \frac{\alpha}{2}\right) dr\\
L_{23}&=+\frac{8 c_v}{\alpha} &&\int^1_0 r \left(\cosh\psi(r)\lambda_\mathrm{ref}^2 \left(\psi(r) - \psi_\mathrm{w} \right)-\frac{\alpha}{2} \sinh\psi(r)\right) dr\\
L_{31}&=-\;\frac{c_v}{\alpha} && \int^1_0 \left(r-r^3 \right)\sinh\psi(r)dr\\
L_{32}&=-\frac{8 c_v}{\alpha} && \int^1_0 r \sinh\psi(r) \left(c_v\int^1_r \frac{1}{r_1}\int^{r_1}_0 r_2 \cosh\psi(r_2) dr_2 dr_1 + \frac{\alpha}{2}\right) dr\\
L_{33}&=-\frac{8 c_v}{\alpha} &\>& \int^1_0 r \left(\sinh\psi(r)\lambda_\mathrm{ref}^2 \left(\psi(r) - \psi_\mathrm{w} \right)-\frac{\alpha}{2}\cosh\psi(r)\right) dr.\\
\end{alignedat}
\end{equation}
We note that (apart from notation) this set of expressions is completely equivalent to the set of equations for $L_{ij}$ by Gross and Osterle \cite{Gross&Osterle}. In a next step, we are concerned with reducing the complexity of the $L_{ij}$ coefficients in Eq. (\ref{eq:Lmatrix}). This can be done first and foremost by reducing the triple integrals to single integrals in $L_{12}, L_{22}$ and $L_{32}$, see \emph{Appendix A}. In the notation of Sasidhar and Ruckenstein \cite{Sasidhar&Ruckenstein} this implies a reduction of the $k_1, k_3$ and $k_7$ integrals, given by
\begin{align}
\begin{split}\label{eq:k1} 
k_1&= \hspace{1.5em}\int^1_0 r\int^1_r \int^{r_1}_0 \frac{r_2}{r_1}\cosh\psi(r_2) dr_2 dr_1 dr = \frac{1}{4}\int^1_0 r \> \left(1-r^2 \right)\cosh\psi(r) dr
\end{split} \\
\begin{split}\label{eq:k3}
k_3 &= \hspace{1.5em}\int^1_0 r \sinh\psi(r)\int^1_r\int^{r_1}_0 \frac{r_2}{r_1}\cosh\psi(r_2) dr_2 dr_1 dr \\ 
&=\hspace{0.5em}-\int^1_0 r \cosh\psi(r)\frac{\lambda^2_\textrm{ref}}{c_v} \left(\psi(r)-\psi_\mathrm{w} \right) dr 
\end{split} \\
\begin{split}\label{eq:k7}
k_7 &= \hspace{1.5em}\int^1_0 r \cosh\psi(r)\int^1_r\int^{r_1}_0 \frac{r_2}{r_1}\cosh\psi(r_2) dr_2 dr_1 dr \\
&=-2 \int^1_0 r\cosh\psi(r)\ln r\left(\frac{1}{2}r^2\cosh\psi(r) - \frac{\lambda^2_\textrm{ref}}{4c_v} \left(r \frac{\partial\psi(r)}{\partial r}\right)^2\right)dr.
\end{split}
\end{align}

In the above equations the reduced form of $k_7$ to a single integral is a new result, and thus by substituting Eqs. (\ref{eq:k1})-(\ref{eq:k7}) into Eq. (\ref{eq:Lmatrix}), we can now show for the first time that all $L_\text{ij}$ expressions can be expressed as single integrals. Computationally this had the advantage that all $L_\text{ij}$ coefficients can be formulated as a first order differential equation in $r$, which is much easier to program and saves computational time.

\section{ Fundamental properties of electrokinetic linear response }

\subsection{ Onsager reciprocal relations } 

With these simplifications, we can now deduce in a straightforward manner that the $L$-matrix must be symmetric. Namely, by substituting Eqs. (\ref{eq:k1}) and (\ref{eq:k3}) into Eq. (\ref{eq:Lmatrix}) it follows that $L_{21}=L_{12}$ and $L_{32} = L_{23}$. Finally, using the boundary conditions of $\psi(x,r)$ and the PB-equation (\ref{eq:PoissonBoltzmanndim}) one can also show that $L_{31}=L_{13}$ and thus prove symmetry of the flux-force matrix.   The final reduced form of the symmetric $L$-matrix can thus be written as
\begin{equation}\label{eq:Lmatrixreduced}
\begin{alignedat}{3}
L_{11}\hspace{1.35em} &= &&\,\frac{1}{8\alpha} &&\\
L_{22} \hspace{1.35em}&= && \frac{8 c^2_v}{\alpha} &
&\enskip k_7 + 4 c_v \int^1_0 r \cosh\psi(r) dr\\
L_{33}\hspace{1.35em} &= &{}-{}&\frac{8 c_v}{\alpha} && \int^1_0 r \left(\sinh\psi(r)\lambda_\mathrm{ref}^2 (\psi(r) - \psi_\mathrm{w})-\frac{\alpha}{2}\cosh\psi(r)\right) dr\\
L_{21}=L_{12}&= &&\;\frac{c_v}{\alpha} && \int^1_0 (r-r^3)\cosh\psi(r)dr \\
L_{31}=L_{13}&=  &&\hspace{0.35em}\frac{4}{\alpha} && \int^1_0 r\lambda^2_\mathrm{ref}(\psi(r) - \psi_\mathrm{w}) dr \\
L_{23}=L_{32}&= &&\frac{8 c_v}{\alpha} && \int^1_0 r \left(\cosh\psi(r)\lambda^2_\mathrm{ref}(\psi(r) - \psi_\mathrm{w}) - \frac{\alpha}{2}\sinh\psi(r)\right)dr
\end{alignedat}
\end{equation}
where the analytic form of $L_{22}$ is new due to the single $k_7$ integral presented in Eq. (\ref{eq:k7}).   

The symmetry of the force-flux linear response matrix, as just shown for Eq. (\ref{eq:fluxesnforces}), is generally known as ``Onsager reciprocity'' or ``Onsager symmetry,'' a phenomenon characteristic of linear response of systems that are near equilibrium. Onsager derived the reciprocal relations for a general thermodynamic force-flux linear response matrix,  based on the assumption that the microscopic equations of motion are reversible \cite{onsager1931,Monroe&Newman}.   Onsager reciprocity is a fundamental postulate of (linear, irreversible) non-equilibrium thermodynamics~\cite{deGroot&Mazur}, which is also assumed in models of electrokinetic phenomena~\cite{hunter_book}, usually without any microscopic justification.  Macroscopic proofs of electrokinetic Onsager reciprocal relations are available for porous media, based on local equilibrium assumptions in formal homogenization theory~\cite{looker2006,schmuck2015}, but we are not aware of explicit proofs based on the microscopic equations of motion for the general situation with salt concentration gradients, as shown here for a cylindrical pore, enabled by our analytical evaluation of the integrals in $k_7$.   In contrast, the classical assumption of constant virtual salt concentration leads to a much simpler $2\times 2$ linear response matrix (e.g., ref.~\cite{YanJPCC}, whose symmetry can be proven for any cross-sectional shape and surface charge distribution~\cite{Bazant_Arxiv}.

\subsection{ Second Law of Thermodynamics }

Any symmetric real matrix has real eigenvalues and orthogonal eigenvectors, but the eigenvalues of the force-flux linear response matrix $L$ must also be positive.  In other words, the matrix must be positive definite.   This property has its roots in the Second Law of Thermodynamics, which states that entropy production is non-negative during an irreversible process.   Using the analytical results above, we are able to prove this property directly from the equations of motion.  

We here define the dissipated power density $P$ in a slice of the cylinder ($x\in\big[a,b]$) as the product of fluxes and conjugate driving forces (i.e., only diagonal elements are used)
\begin{equation}
P = -\overline{u_x \yPh}\cdot\frac{\Delta p_{t,v}(x)}{b-a} -  \overline{j_{\mathrm{ions},x}}\cdot\frac{\Delta \mu_v(x)}{b-a}- \overline{j_{\mathrm{ch},x}}\cdot\frac{\Delta \phi_v(x)}{b-a}
\end{equation}
which is analogous to the definition in~\cite{Bazant}. If we were to re-assign dimensions to this equation we would see that it is a power density with units of W/m$^{3}$. 
By the Second Law of Thermodynamics, this equation has to be positive as the process it describes is irreversible. Now, passing to the limit $a \to b$ we see that  
\begin{equation}\label{eq:entropygeneration}
P = \overline{u_x \yPh}\cdot\left(-\frac{\partial p_{t,v}}{\partial x}\right) +  \overline{j_{\mathrm{ions},x}}\cdot\left(-\frac{\partial \mu_v}{\partial x}\right) + \overline{j_{\mathrm{ch},x}}\cdot\left(-\frac{\partial \phi_v}{\partial x}\right).
\end{equation}
Finally, we observe that when we insert Eq. (\ref{eq:fluxesnforces}), we can write Eq. (\ref{eq:entropygeneration}) as 
\begin{equation}
\left(-\frac{\partial p_{t,v}}{\partial x},-\frac{\partial \mu_v}{\partial x},-\frac{\partial \phi_v}{\partial x}\right)\cdot L\cdot \left(-\frac{\partial p_{t,v}}{\partial x},-\frac{\partial \mu_v}{\partial x},-\frac{\partial \phi_v}{\partial x}\right)^t >0
\end{equation}
which is 
a statement of positive definiteness of the matrix $L$, because it should hold for arbitrary driving forces. 

\subsection{ Change of basis }

In the last part of this section, we analyze the flux-force matrix formalism more generally and come to the conclusion that there are many possible (actually, an infinite number of) 
coupled sets of flux-force equations equivalent to the set in Eq. (\ref{eq:Lmatrix}) in the sense that 
Onsager symmetry 
is preserved
and the dissipation rate is described by the product of fluxes and conjugate forces (while there is also an infinite set of relationships that does not have Onsager symmetry).
Gross and Osterle \cite{Gross&Osterle} already showed quite extensively the equivalence of Eq. (\ref{eq:Lmatrix}) and a coupled set with $\overline{u_x \yPh},\overline{j_{\mathrm{diff},x}},\overline{j_{\mathrm{ch},x}}$ as fluxes and $-\partial_x p_{h,v},-\partial_x\pi_v,-\partial_x\phi_v$ as driving forces. Here, differential flow is defined as 
\begin{equation} \label{eq:j_diff}
 \bm{j}_\mathrm{diff}(x,r) = \frac{\bm{j}_\mathrm{ions}(x,r)}{2c_v(x)} - \bm{u}(x,r).
\end{equation}
However, it is quite an arduous effort to verify the claims in ref. \cite{Gross&Osterle}, as the authors performed the change of coupled relations simultaneously with the reduction of the integrals in Eq. (\ref{eq:Lmatrix}). Interestingly, we found that their specific claims can be formulated in terms of a much more general case, very similar to the one described by de Groot and Mazur \cite{deGroot&Mazur} as we will outline next. Let $\bm{J}$ denote a set of fluxes in the $x$-direction and $\bm{X}$ a set of coupled thermodynamic forces, such as $\bm{J}=\left(\overline{u_x \yPh},\overline{j_{\mathrm{ions},x}},\overline{j_{\mathrm{ch},x}}\right)$ and $\bm{X}=\left(-\partial_x p_{t,v},-\partial_x \mu_v,-\partial_x \phi_v\right)$. Let $\bm{J'}$ and $\bm{X'}$ be another coupled set of fluxes and driving forces, so that we have the relations
\begin{equation}\label{eq:Lrelation}
\bm{J} = L\cdot \bm{X} \quad\text{and}\quad \bm{J'}= L' \cdot\bm{X'}
\end{equation}
and the dissipation rate can be written in this notation as
\begin{equation}
P = \bm{J} \cdot \bm{X} = \bm{X}^t\cdot L\cdot \bm{X}. 
\end{equation}
Let us also define the (invertible) linear maps $A: \bm{R}^3 \rightarrow \bm{R}^3$ and $B: \bm{R}^3 \rightarrow \bm{R}^3$ by the relations
\begin{equation}
\bm{J'}=A\cdot\bm{J} \quad\text{and}\quad \bm{X'} = B\cdot\bm{X}
\end{equation}
making them the transformations 
that carry $\bm{J}$ onto $\bm{J'}$ and $\bm{X}$ onto $\bm{X'}$. In general one can easily deduce that the equation
\begin{equation}
L' = A\cdot L \cdot B^{-1}
\end{equation}
describes the relation between the coupled flux-force equations.
Assuming that these transformations 
are non-trivial (thus invertible) one can quite easily prove conservation of Onsager symmetry and invariance of the dissipation rate under the associated change of basis
 (for arbitrary $\bm{X}\in\bm{R}^3)$ if 
\begin{equation}
A^t = B^{-1}.
\end{equation}
Indeed, if this relation is assumed to hold we observe
\begin{equation}
(L')^t = (A\cdot L \cdot B^{-1})^t = (B^{-1})^t \cdot L^t \cdot A^t = A \cdot L \cdot B^{-1}  
\end{equation}
and
\begin{equation}
P' = \bm{J'}\cdot \bm{X'} = A\cdot \bm{J} \cdot B \cdot \bm{X} = \bm{J} \cdot A^t \cdot B \cdot \bm{X} =\bm{J}\cdot \bm{X} = P
\end{equation}
which we set out to show.

We note that if we work with $\bm{J'} = \left(\overline{u_x \yPh},\overline{j_{\mathrm{diff},x}},\overline{j_{\mathrm{ch},x}}\right)$ and $\bm{X'} = \left(-\partial_x p_{h,v},-\partial_x \pi_v,-\partial_x \phi_v\right)$ and our original sets, then A and B are given by
\begin{equation}
A=\begin{pmatrix}1 & 0 & 0\\  -1 & \frac{1}{2 c_v} & 0\\ 0 & 0 & 1 \end{pmatrix}\quad ,\quad B=\begin{pmatrix}1 & 2c_v & 0\\  0 & 2 c_v & 0\\ 0 & 0 & 1 \end{pmatrix}.
\end{equation}
It is now straightforward to verify that $A^t = B^{-1}$, thereby proving the claim of Gross and Osterle \cite{Gross&Osterle}. The matrix $L'$ can be calculated directly using these expressions and Eq. (\ref{eq:Lrelation}) and was found to be in agreement with Eq. (22) in Ref. \cite{Gross&Osterle}.


\section{Uniform Potential model}
For pores that are thin relative to the Debye length, concentration profiles across the pore are only weakly changing and we can simplify the above framework significantly~\cite{Bocquet&Charlaix}. This simplification goes under various names, such as ``fine capillary pore model'', ``uniform potential (UP) model'' \cite{Schlogl, Verbrugge&Hill, Bowen&Welfoot, Sonin, HawkinsCwirko&Carbonell, kim_microfluid, Biesheuvel}, and also as Teorell-Meyers-Sievers (TMS) theory, though TMS-theory does not include fluid flow \cite{Teorell,Tedesco}. 
In the UP-model, the coefficient-matrix $L$ of Eq. (\ref{eq:Lmatrixreduced}) simplifies to
\begin{equation}\label{eq:LmatrixUP}
\begin{alignedat}{4}
L_{11}\hspace{1.35em} &=&& && \frac{1}{8\alpha} &&\\
L_{22}\hspace{1.35em} &=  &&2 \> c_v \cosh\psi &{}+{}&\frac{c_v^2}{2\alpha}&\>\cosh&^2\psi \\
L_{33}\hspace{1.35em} &=  &&2 \>c_v \cosh\psi &{}+{}&\frac{c_v^2}{2\alpha}&\sinh&^2\psi\\
L_{21}=L_{12} &= &&&{}+{}&\frac{c_v}{4\alpha}&\cosh&\hphantom{^2}\psi \\
L_{31}=L_{13} &= &&& {}-{}&\frac{c_v}{4\alpha}&\sinh&\hphantom{^2}\psi \\
L_{23}=L_{32} &= &-&2 \> c_v \sinh\psi &{}-{}&\frac{c_v^2}{2\alpha}&\sinh&\hphantom{^2}\psi\cosh\psi
\end{alignedat}
\end{equation}
which is now independent of the exact pore geometry, except for a factor $\alpha$, originally based on the geometry of a capillary pore, but adjustable to describe other pore geometries.

To simplify Eq. (\ref{eq:LmatrixUP}) we use Eq. (\ref{eq:Boltzmanndim}) to derive
\begin{align}\label{eq:donnan}
\begin{split}
\omega X&=c_- - c_+ = 2\text{ }c_v\sinh\psi \\
c_\textrm{T}&=c_- + c_+=2\text{ }c_v\cosh\psi=\sqrt{X^2+(2c_v)^2}
\end{split}
\end{align}
where $X$ is the magnitude of the density of fixed charges in the nanopore, defined as number of charges per unit pore volume, taken as a positive number (unrelated to $\bm{X}$ of the previous section), while $\omega$ is the sign of the membrane charge (e.g., $\omega$=+1 for a nanopore or membrane with fixed positive charges, i.e., an anion-exchange membrane). Furthermore, $c_\textrm{T}$ is the total ions concentration in the pore, which is always larger than $X$, see Eq. (\ref{eq:donnan}b). Inserting Eq. (\ref{eq:donnan}) in Eq. (\ref{eq:LmatrixUP}) results for the coefficients of the $L$-matrix in
\begin{equation}\label{eq:LmatrixUP_simplified}
\begin{alignedat}{2}
L_{11}\hspace{1.35em} &=  &&\frac{1}{8\alpha}\\
L_{22}\hspace{1.35em} &=  &&L_{11}\>c_\textrm{T}^2+c_\textrm{T} \\
L_{33}\hspace{1.35em} &=  &&L_{11}\> X^2 +c_\textrm{T}\\
L_{21}=L_{12} &= &&L_{11}\>c_\textrm{T} \\
L_{31}=L_{13} &=  &{}-{}&L_{11}\>\omega X \\
L_{23}=L_{32} &= &{}-{}&L_{11}\>\omega X c_\textrm{T} -\omega X
\end{alignedat}
\end{equation}
of which the determinant can now easily be derived to be
\begin{equation*}
D=L_{11} \left(c_\textrm{T}^2-X^2\right)=L_{11}\left(2c_v\right)^2
\end{equation*}
which is strictly positive.

Above we have now given the coefficients of the $L$-matrix for the UP-model where forces are gradients in virtual quanties $p_{t,v}$, $\mu_v$, and $\phi_v$ as in Eq. (\ref{eq:fluxesnforces}). However, the model can be further simplified when we return to ``real'' pressures, concentrations and potentials. The resulting set of equations is
\def\dPdx{\vphantom{\frac{\partial p^h}{\partial x}}}
\begin{equation}\label{eq:UP}
\begin{alignedat}{3}
u \hspace{0.85em}&= & {}-{}L_{11}\left(\dPdx\right.&\frac{\partial p^h}{\partial x}&{}-{}\omega X\frac{\partial \phi}{\partial x} & \left.\dPdx\right)\\
j_\textrm{ions}\hspace{0.35em} &= &c_\textrm{T} u {}-{}&\frac{\partial c_\textrm{T}}{\partial x}&{}+{}\omega X\frac{\partial \phi}{\partial x}&\\
j_\textrm{ch} &= &{}-{}\omega X u {}+{}\omega&\frac{\partial X}{\partial x}&{}-{}c_\textrm{T} \frac{\partial \phi}{\partial x}&
\end{alignedat}
\end{equation}
where we neglect overbar signs to denote pore-averaged fluxes. For a constant membrane charge, $X$, the term $\omega\partial_x X$ is zero. At the two pore mouths (on either side of the pore) we have to solve for step changes across the EDLs at the membrane/solution interfaces, leading to jumps in $p^h$, $c_\textrm{T}$ and $\phi$, using the Donnan (Boltzmann) relations
\begin{align}\label{eq:donnan_tms}
\begin{split}
p^h_m\hspace{0.35em}&=p^h_\textrm{ext}+c_{\textrm{T},m}-2c_\textrm{ext}\\
c_{\textrm{T},m}&=\sqrt{X^2+\left(2c_\textrm{ext}\right)^2}\\
\phi_m\hspace{0.35em}&=\phi_\textrm{ext}+\sinh^{-1}{\omega X / 2 c_\textrm{ext}}
\end{split}
\end{align}
where subscript ``$m$'' refers to a position just within the membrane, beyond the EDLs at the membrane/solution interface and where ``ext'' describes a position just outside the membrane, in the electroneutral electrolyte. 

\section{Results and Discussion}

\subsection{ Numerical solution }

Although the capillary pore model assumes local quasi-equilibrium, which implies local linear response, axial variations lead to global nonlinear response of the charged nanopore, which can be far from equilibrium. Therefore the capillary pore model must be solved numerically along the length of the pore ($x$-direction), to find the profiles of the virtual quantities $p_{t,v}(x)$, $\mu_v(x)$ and $\phi_v(x)$. 
Because the pore-averaged fluxes, $\overline{u_x \yPh}$, $\overline{j_{\mathrm{ions},x}}$ and $\overline{j_{\mathrm{ch},x}}$, are invariant along the pore, this calculation requires solving a system of three first order, quasi-linear ordinary differential equations (ODEs), since the Onsager matrix of Eq. (\ref{eq:fluxesnforces}) depends on the virtual fields. The six cross-coefficients $L_\text{ij}$ only depend on wall charge, pore radius, and local (virtual) concentration $c_v$ and thus for a certain charge and radius, can be calculated a-priori as function of $c_v$, and the result stored as six polynomial functions of $L_\text{ij}$ versus $c_v$ and used in the solution of the three ODEs in which coordinate $x$ is the running parameter. In this a-priori calculation the PB-equation is solved in radial direction and the profile of $\psi(r)$ calculated, see Fig. 2. After the functions $L_\text{ij}(c_v)$ have been determined, the PB-calculation based on $\sigma$ and $\psi(r)$ is no longer necessary.

Next, using the expressions given in \emph{Appendix B}, we evaluate the $r$-dependence of the $x$-directional fluxes. Note that for these fluxes there is no Onsager symmetry for the flux-force framework, and thus all nine cross-coefficients must be separately analyzed. After that, using the continuity equation (\ref{eq:continuity}), we solve for the radial components of the fluxes, as their axial component is known on every point of the grid, which reduces the continuity equation to a first order $r$-dependent differential equation, to generate streamline- and vectorfield-plots for $\bm{u}$, $\bm{j}_{\mathrm{ions}}$, and $\bm{j}_{\mathrm{ch}}$.
We illustrate the capillary pore model with the example of energy harvesting from salinity differences by flows through charged nanopores.

\subsection{Pore-averaged fluxes}
Calculations presented in this section are based on a pore placed between two electrolyte solutions with $c_\mathrm{ext}=500$ and 10 mM salt concentration. We use a pore radius of $R=2$ nm, pore length of $L=100$ $\mu$m, an average ion diffusion coefficient of $D=2\cdot10^{-9}$ m$^2$/s, viscosity of $\mu=1$ mPa.s, and temperature $T=298$~K. The permittivity of water is $\varepsilon=6.91\cdot 10^{-10}$ F/m, thermal voltage is $\Phi_\text{B}=25.7$ mV, and surface charge is $\sigma = -10\text{ mC/m}^2$. Thus we have $\lambda_\mathrm{ref}=4.79$ and $\alpha=202$. We assume the two reservoirs to have the same hydrostatic pressure, thus $p_{h,v}(x=0) = p_{h,v}(x=1)$. We apply a current of $\overline{j_{\mathrm{ch},x}} = 140$ which translates dimensionally to 27 mA/cm$^2$. 

\begin{figure}[H]
\centering
\includegraphics[scale=0.55]{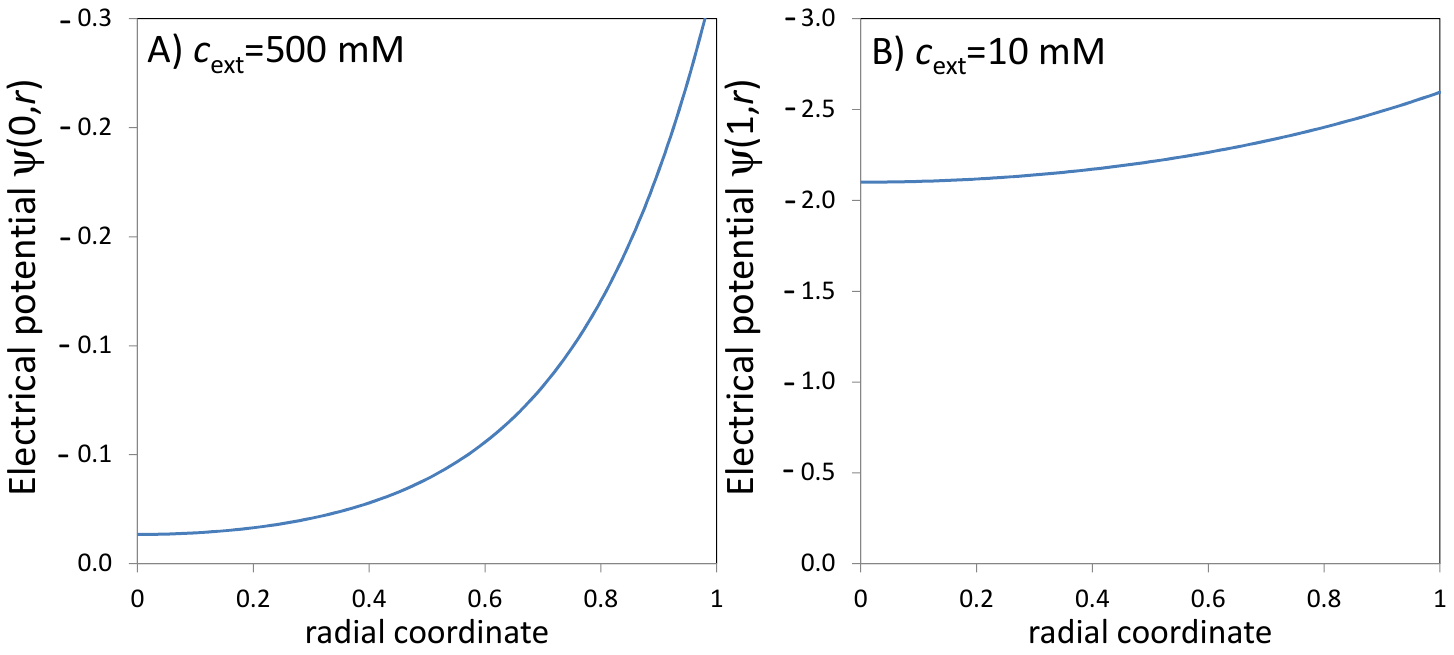}
\caption{Dimensionless electric potential $\psi(0,r)$ and $\psi(1,r)$ in a charged nanopore ($R= 2$ nm; surface charge $\sigma=-10$ mC/m$^2$) at the two ends of the pore, as a function of radial coordinate. 
}\label{fig:Potentiaalplotjes}
\end{figure}

For external salt concentrations of $c_\mathrm{ext}$=500 mM (at the left-hand pore entrance, where $x=0$) and $c_\mathrm{ext}$=10 mM (at the right-hand side, $x=1$), we can directly calculate the potential profile $\psi(x,r)$, as plotted in Fig. \ref{fig:Potentiaalplotjes}. Because the Debye length increases through the pore due to its reciprocal dependence on $c_v(x)$, we see that at at $x=0$ (panel A) $\psi(x,r)$ drops off faster (relatively) from the wall towards the pore axis than at $x=1$ (panel B). Also, the magnitude of $\psi(x,r)$ is much larger at $x=1$ for $c_\mathrm{ext}$=10 mM [Note that scales in panels A) and B) are different].

\begin{figure}[H]
\centering
\includegraphics[scale=0.55]{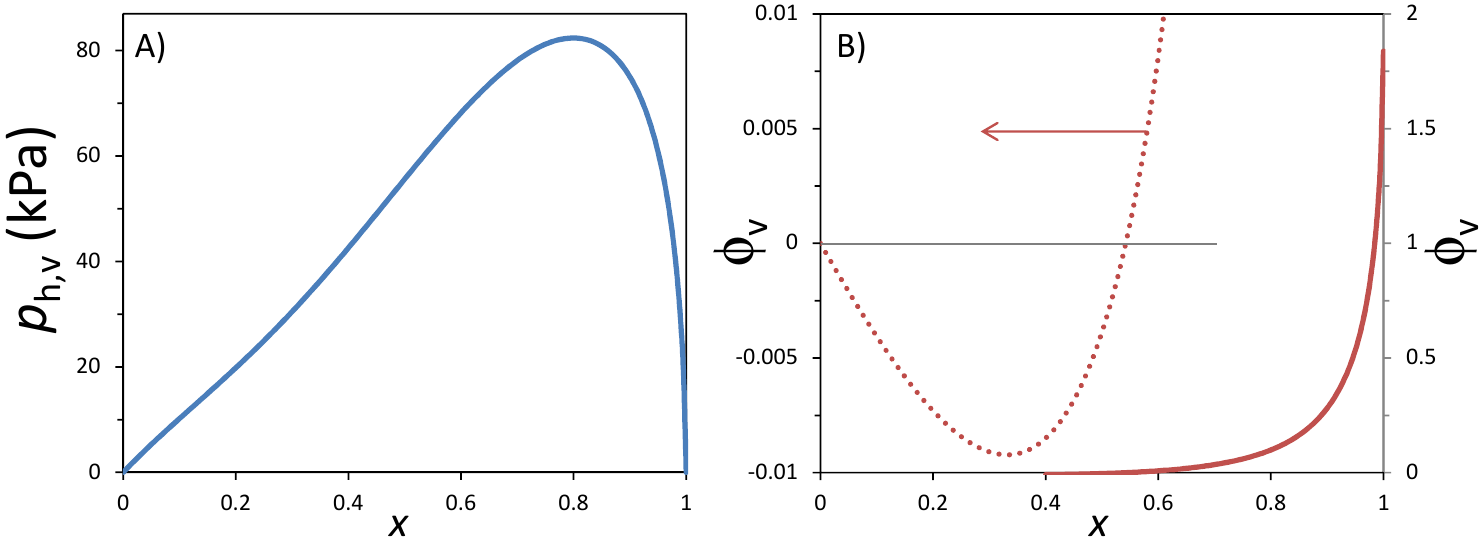}
\caption{Plots of A) virtual hydrostatic pressure $p_{h,v}(x)$, and B) virtual electric potential, $\phi_v(x)$, versus axial position in pore, $x$. In B) the dashed line is an enlarged view of the potential profile for $x<$0.6.}
\label{fig:Phphiplot}
\end{figure}

For the virtual quantity $c_v(x)$ (and thus $\mu_v(x)=\ln c_v(x)$ and $p_{t,v}(x)$) we find a gradual change from one end of the pore to the other [not shown]. However, for virtual hydrostatic pressure $p_{h,v}(x)$ and virtual electric potential $\phi_v(x)$, the behavior is more interesting, see Fig. \ref{fig:Phphiplot}. First of all, hydrostatic pressure, though zero at both pore mouths, makes a steep excursion within the pore, as also observed in ref.~\cite{Holt}, reaching a maximum value of $p_{h,v}\sim82$ kPa, corresponding to the osmotic pressure of a 17 mM salt solution. The electric potential, $\phi_v(x)$, is virtually unchanging for most of the pore ($0<x<0.8$) before steeply increasing at the very end of the pore. Interestingly, $\phi_v(x)$ is slightly negative at the beginning of the pore before turning positive. Note that in Fig. \ref{fig:Phphiplot} virtual quantities are discussed, not ``real'' pressures or potentials.

Concerning fluxes, the average flux of ions is $\overline{j_{\mathrm{ions},x}}=18.7$ mmol/m$^2$/s and the average velocity of fluid is $\overline{u_{x}}=-0.49$ $\mu$m/s for the chosen parameter set. This implies that ions move right, while the water flows left, in agreement with the common notion of solvent flowing to the side of lower total pressure (the side of higher salinity in case of equal hydrostatic pressure).

\subsection{Analysis of fluxes as function of $r$-coordinate}
For the $r$-dependence of the $x$-component of the fluxes, we find for water velocity $u_{x}(r)$ an almost parabolic shape (with no-slip at the wall), essentially unvarying from $x=0$ to $x=1$. Ion flux $j_{\mathrm{ions},x}(r)$ does not vary much with axial nor radial coordinate, from a value of $\sim 17.4$ mmol/m$^2$/s in the center, to $\sim 19.6$ mmol/m$^2$/s at the wall [not shown]. Thus note that the highest ion fluxes are found at the wall. 

\begin{figure}[ht]
\centering
\subfigure{
  \includegraphics[width=7cm]{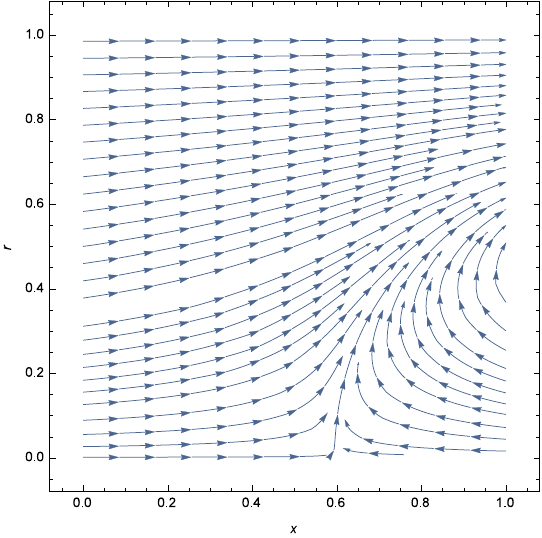}}
\quad
\subfigure{%
  \includegraphics[width=7cm]{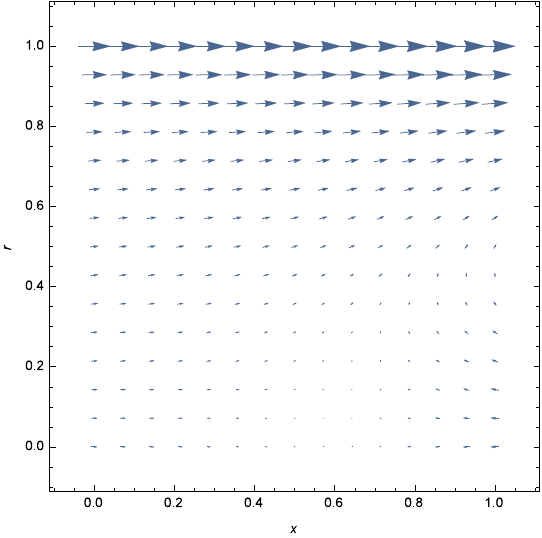}}
\caption{Streamline (left) and vectorfield (right) plots of ionic current $\bm{j}_\mathrm{ch}(x,r)$. The streamline plot clearly demonstrates the inversion of $\bm{j}_\mathrm{ch}(x,r)$, while the vectorfield plot shows the higher magnitude of ionic current near the pore wall.}
\label{fig:jcharge}
\end{figure}

For the profile in ionic current, $j_{\mathrm{ch},x}(r)$, we also find the highest value at the wall, but interestingly, in the center of the pore, the ionic current inverts. In particular, at $x=0$, $j_{\mathrm{ch},x}(r)$ is always positive, increasing from $j_{\mathrm{ch},x}(0)=12$ mA/cm$^2$ to $j_{\mathrm{ch},x}(1)=58$ mA/cm$^2$. However, at $x=1$, $j_{\mathrm{ch},x}(0)=-16$ mA/cm$^2$ while $j_{\mathrm{ch},x}(1)=73$ mA/cm$^2$. This change-over in $j_{\mathrm{ch},x}(0)$ from positive at $x=0$ to negative at $x=1$, implies that there is a ``surface'' within the pore where the $x$-component of the ionic current is zero, as indeed shown in Fig. \ref{fig:jcharge}. As Fig. \ref{fig:jcharge} illustrates, even though the average ionic current is positive (directed to the right), in a range of $r$-positions around the center axis ionic current enters the pore on the right-hand side and flowing to the left, before looping back and exiting the pore again on the right, but now closer to the wall.

In presenting streamline and vectorfield plots in Fig. \ref{fig:jcharge}, one might notice a paradox, as we previously assumed equilibrium in $r$-direction, which should result in $j_{\mathrm{ch},r}=0$ for all $x,r$.  This seems to clash with our calculation of $\bm{j}_\mathrm{ch}$ by virtue of the continuity equation (\ref{eq:continuity}), which very clearly results in a vector field of $\bm{j}_\mathrm{ch}(x,r)$ that has non-zero $r$-components. This paradox is solved by noticing that we normalized our $x$- and $r$-coordinates to a $\big[0,1]\times\big[0,1]$ square. To obtain the true magnitude of our vector components one has to multiply all $r$-components by $R$ and all $x$-components by $L$. The latter is much larger and this justifies the claim that the $r$-component of $\bm{j}_\mathrm{ch}(x,r)$ is almost $0$. 

To our knowledge this is the first time that for a long and narrow charged pore, computations of the capillary pore model are made in the presence of an axial concentration gradient. Ref. \cite{WymanCostin} considered an axial concentration gradient but their method involved solving the NS-, NP- and PB-equations directly, for a system far from the ``needle limit'' of  $L/R\rightarrow\infty$. Instead, the geometry considered was for a pore even wider than long (i.e., $L/R<1$). In ref. \cite{WymanCostin} an inversion within the pore of one of the fluxes was observed, namely in $\bm{u}_x(x,r)$. In ref. \cite{Sounart&Baygents} electrically drived fluid vortices were predicted in microchannels in ampholytic salt solutions. Ref.~\cite{Kang} modelled in two dimensions the full problem of transport in a cylindrical pore between two solutions of different salt concentration, while ref.~\cite{Yeh2014} solved the problem for a conical nanopore in the absence of fluid flow. Our analysis, therefore, provides a new perspective on the generality of this phenomenon. We hope that calculating the full vector fields of $\bm{u}(x,r), \bm{j}_\mathrm{ions}(x,r)$ and $\bm{j}_\mathrm{ch}(x,r)$ via the formulation of averaged fluxes will prove useful to find other flux inversions as well.

\subsection{Energy generation from a salinity difference}
Next, we show how our calculations can provide relevant information on the performance of an electrokinetic energy harvesting device based on a salt concentration difference. Here, we consider the single membrane pore as part of a membrane which is placed in a stack of multiple membranes, with alternating sign of the fixed charge on the membrane. This process is called reverse electrodialysis \cite{Fair&Osterle,post2008,dlugolecki2009,Siria,Rankin,Tedesco}. Because of the salt concentration difference across each membrane, power is delivered to a load $R$ placed in an external circuit, see Fig. \ref{fig:Poremodel}. In the remainder of this section, for average, axial, fluxes we drop the overbar-sign, and we also drop subscript ``$x$''.

\begin{figure}[H]
\centering
\includegraphics[scale=0.35]{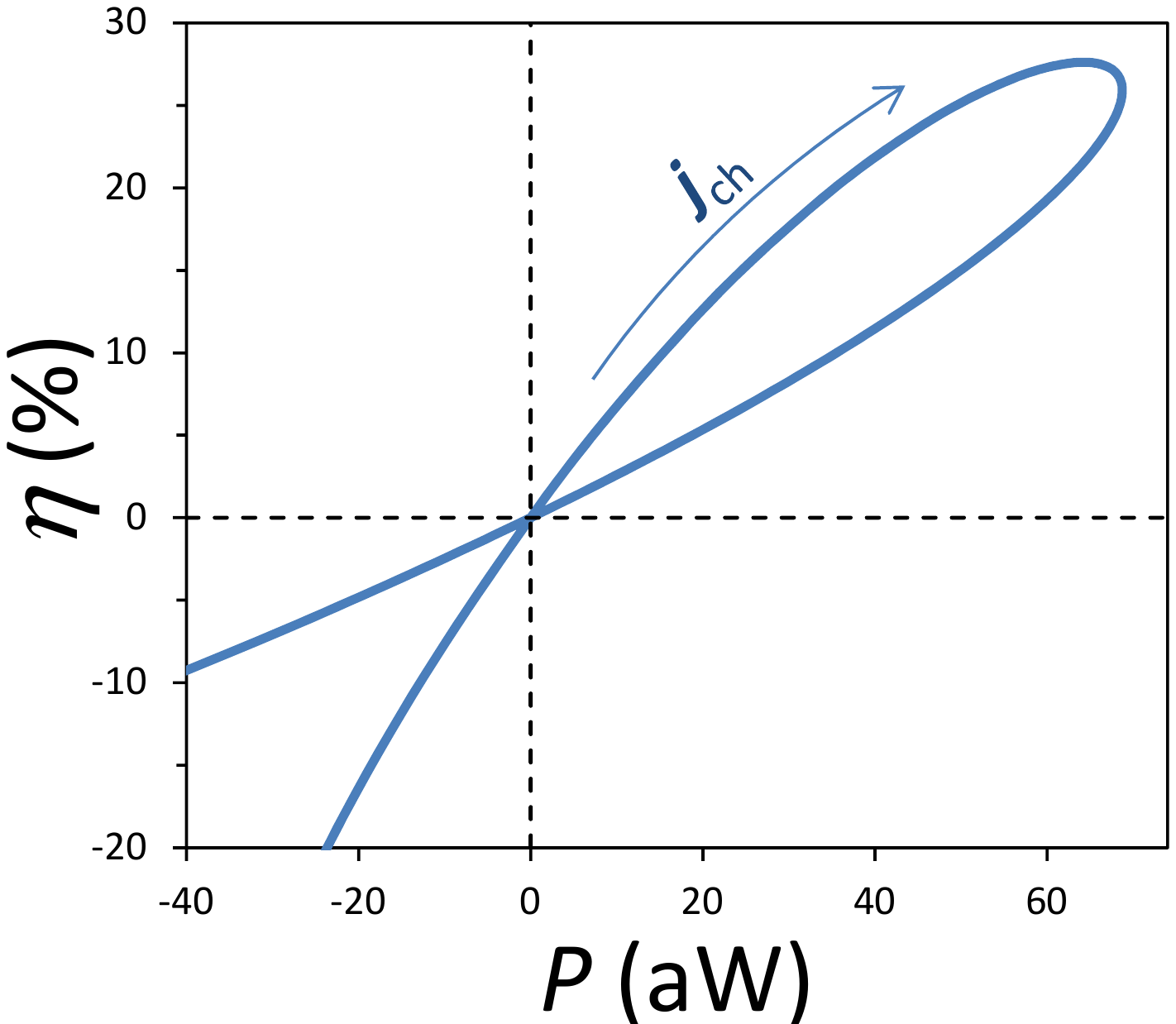}
\caption{Plot of energy efficiency $\eta$ against generated  electrical power $P$ by a single charged nanopore placed between two electrolytes of different salinity ($c_\mathrm{ext}=10$ and 100 mM). Current increases in direction of arrow.}\label{fig:powerplot}
\end{figure}

We can define a local efficiency of the generation of electrical energy, at any point in the membrane pore, as 
\begin{equation} \label{eq:efficiency}
\quad\eta' = 
-\frac{{j_{\mathrm{ch}}}\partial_x \phi_v}{{j_{\mathrm{ions}}}\partial_x \mu_v+{u_{}}\partial_x p_{t,v}}.
\end{equation}
For a zero overall hydrostatic pressure, separately integrating the upper and lower side of Eq. (\ref{eq:efficiency}) over the entire pore length, results for the overall energy efficiency $\eta$ in~\cite{Fair&Osterle,Yeh2014}
\begin{equation} \label{eq:efficiency_II}
\eta = -\frac{{j_{\mathrm{ch}}}\Delta \phi_v}{{j_{\mathrm{ions}}}\Delta\mu_v-{u}\Delta \pi_v}
\end{equation}
noting that differences $\Delta$ are defined as conditions at ``$x=1$'' minus at ``$x=0$''. In Fig. \ref{fig:powerplot} we plot $\eta$ versus generated electrical power by a single nanopore (instead of plotting both versus current or voltage as in ref. \cite{Tedesco}). These calculations are based on external salt concentrations of $c_\mathrm{ext}=100$ and 10 mM, with all other parameter settings the same as before. The maximum in energy conversion efficiency of $\eta\sim28\%$ is obtained for a current of $\sim 19$ mA/cm$^2$ ($\Delta \phi_v \sim 31$ mV; salt transport efficiency $\vartheta=j_\mathrm{ch}/j_\mathrm{ions} \sim 63\%)$. Around this optimum, the fluid velocity, $u$, switches from normal osmosis (directed to the high-salinity side at lower currents), to anomalous osmosis at higher currents where fluid flows to the low-salinity side \cite{Sasidhar&Ruckenstein_II}. In the present calculation, fluid velocity is found to change from $u=-0.7$ $\mu$m/s for open-circuit conditions (zero current, $\Delta \phi_v \sim 47$ mV) to $u=+1.6$ $\mu$m/s for electrical short-circuit ($\Delta \phi_v=0$, current $\sim 54$ mA/cm$^2$). Values for the power per pore in aW as depicted in Fig. \ref{fig:powerplot} can be multiplied by 80 to a power in mW/m$^2$ pore area, resulting in a maximum power of $\sim 0.64$ W/m$^2$.

\subsection{Calculations using Uniform Potential (UP) model}
Finally, we applied the UP-model and made the same calculations as before. Comparing with results in Fig. \ref{fig:Potentiaalplotjes}, we calculate for the $r$-independent $\psi$-value in the UP-model, that $\psi_{c_\textrm{ext}=500\textrm{ mM}}=-0.10$ and $\psi_{c_\textrm{ext}=10\textrm{ mM}}=-2.35$, values in between minimum and maximum $\psi(r)$-values in Fig. \ref{fig:Potentiaalplotjes}. Plots of virtual hydrostatic pressure and potential in Fig.~\ref{fig:Phphiplot} come out almost exactly the same with the UP-model, with the maximum in $p_{h,v}$ somewhat higher at $p_{h,v}\sim 100$ kPa, reached at a slightly higher $x\sim 0.85$. For potential $\phi_v$, again an initial decay is predicted, with $\phi_v$ turning positive at $x\sim0.56$ to end at $x=1$ at $\phi_v=1.877$, which is $\sim2$ \% above the result of the full capillary model. For efficiency vs. power, as in Fig. \ref{fig:powerplot}, results match almost exactly, with the maximum in efficiency $\eta_\textrm{max}$ for both models at $j_\textrm{ch}\sim 20$ mA/cm$^2$, with $\eta_\textrm{max} \sim 27.6$ \% for the full model and 27.9 \% for the UP-model. Fluid flow $u$ in both models switches sign just below $j_\textrm{ch}$=20 mA/cm$^2$, and increases with $j_\textrm{ch}$. However, water velocity increases somewhat faster for the UP-model: at $j_\textrm{ch}=39$ mA/cm$^2$, $u=0.77$ $\mu$m/s for the full model and $u=0.93$ $\mu$m/s for the UP-model. 

\newpage

In conclusion, the UP-model (TMS-model, fine capillary pore model) gives predictions for the overall (pore-averaged) transport in thin capillary pores which are in almost quantitative agreement with the full model, even for conditions where the Debye length is about the pore size on one end of the pore and much smaller than the pore size on the other end. For larger pores, the UP-model is expected to deviate more significantly. Furthermore, the UP-model does not provide information on microscopic phenomena such as the development of loops in current or fluid flow. Also, calculations [not reported here] show that the UP-model can significantly overpredict co-ion exclusion (i.e., the full capillary pore model predicts a significantly higher pore-averaged concentration of co-ions).
 
\section{Conclusions}
We have analyzed the capillary pore model, which is a semi-analytical model of ion transport and flow through charged nanopores, based on the assumption of local quasi-equilibrium, allowing for overlapping EDLs and axial concentration gradients. The analysis is based on the force-flux framework of Osterle and coworkers \cite{Gross&Osterle,Fair&Osterle}, for which we have discovered a simple single-integral expression for the coefficient $k_7$. We demonstrate that all symmetric force-flux frameworks are equivalent and obey Onsager reciprocal relations for local linear response, as a result of the local quasi-equilibrium assumption. We also solve the full nonlinear model numerically without integrating over the cross section to resolve the axisymmetric two-dimensional profiles of ion transport and flow. Calculations for a pore subjected to two reservoirs with different salt concentrations, as a model of reverse electrodialysis, demonstrate how in the presence of an overall concentration difference a ``current loop'' can develop at one of the pore ends. We present a plot of energy efficiency versus electrical power generated by a single charged nanopore in this process. We analyze the uniform potential model (fine capillary pore model), a model in which potential and concentration are assumed to be invariant with radial coordinate, and show that for the parameter range investigated, it gives predictions of macroscopic transport properties that are in line with results of the full capillary pore model. Our work unifies previous theoretical work and provides a rigorous basis for further modeling of transport in charged membranes and nanopores. 

\section*{Acknowledgments} 
This work is part of the Delta-ITP consortium, a program of the Netherlands Organisation for Scientific Research (NWO) that is funded by the Dutch Ministry of Education, Culture, and Science (OCW). MZB acknowledges partial support from the Global Climate and Energy Project at Stanford University and by the US Department of Energy, Basic Energy Sciences through the SUNCAT Center for Interface Science and Catalysis. This work was performed in the cooperation framework of Wetsus, European Centre of Excellence for Sustainable Water Technology (www.wetsus.eu). Wetsus is co-funded by the Dutch Ministry of Economic Affairs and Ministry of Infrastructure and Environment, the Province of Frysl\^an, and the Northern Netherlands Provinces.

\newpage

\section*{Appendix A. Derivation of $k$-integrals}

In Appendix A we show how the triple integrals for $k_1, k_3$ and $k_7$ can be reduced to single integrals. For $k_1$ and $k_3$ this allows us to show Onsager Symmetry of the flux-force framework. Another advantage is that single integrals are numerically much easier to evaluate. It was already known to Sasidhar and Ruckenstein\cite{Sasidhar&Ruckenstein} that several of the integrals can be reduced to single integrals. Gross and Osterle\cite{Gross&Osterle} also reduced their expressions to simple forms. We follow the definitions of Sasidhar and Ruckenstein\cite{Sasidhar&Ruckenstein} to define $k_1, k_3$ and $k_7$ as
\begin{align}
k_1 &= \int^1_0 r \int^1_r \frac{1}{r_1} \int^{r_1}_0 r_2 \cosh\psi(r_2) dr_2 dr_1 dr\\
k_3 &= \int^1_0 r \sinh\psi \int^1_r \frac{1}{r_1} \int^{r_1}_0 r_2 \cosh\psi(r_2) dr_2 dr_1 dr\\
k_7 &= \int^1_0 r \cosh\psi\int^1_r \frac{1}{r_1} \int^{r_1}_0 r_2 \cosh\psi(r_2) dr_2 dr_1 dr 
\end{align}
which appear in the matrix elements $ L_{12}, L_{32}$ and $L_{22}$, and in the full calculation are $x$-dependent.

We show that these triple integrals can be reduced, by dividing the area of integration in a suitable way, and switching the order of integration. 

First of all, we note that in general it holds that
\begin{equation}\label{eq:integraaltruc_appendix}
\int^1_r  \left(\int^{r_1}_0 dr_2 \right)dr_1= \int^r_0 \left( \int^1_{r} dr_1 \right)dr_2+ \int^1_r \left( \int^1_{r_2} dr_1 \right) dr_2
\end{equation}
valid because \{$r\leq r_1\leq 1$ and $0\leq r_2\leq r_1$\} is equivalent to the statement \{$\big[0,r]$, $r\leq r_1 \leq 1$ and on $\big[r,1]$, $r_2\leq r_1\leq 1$\}. In Eq. (\ref{eq:integraaltruc_appendix}) on the left-hand side the integration is performed first over $dr_2$ and then over $dr_1$, and on the right-hand side this order is reversed. Thus, we combine Eq. (\ref{eq:integraaltruc_appendix}) with changing the order of integration twice (we first switch $r_1$ and $r_2$, and then $r_2$ and $r$), with the Poisson-Boltzmann equation (in the case of $k_3$ and $k_7$), with partial integration, and with a symmetry argument in the case of $k_7$. We start by reducing the $k_1$ integral according to
\begin{equation}
\begin{alignedat}{2}
k_1 &= &&\int^1_0 r \left( \int^1_r \int^{r_1}_0 \frac{r_2}{r_1}\cosh\psi(r_2) dr_2 dr_1 \right) dr \\
&= &&\int^1_0 r\left(\int^r_0\int^1_r \frac{r_2}{r_1}\cosh\psi(r_2) dr_1 dr_2 + \int^1_r\int^1_{r_2} \frac{r_2}{r_1}\cosh\psi(r_2) dr_1 dr_2\right) dr\\
&= &&\int^1_0 r\left(\int^r_0\ln r_1 |^1_r \textrm{ }r_2\cosh\psi(r_2) dr_2 + \int^1_r r_2\ln r_1 |^1_{r_2} \cosh\psi(r_2) dr_2\right)dr\\
&=  {}-{}&&\int^1_0 r\left(\ln r\int^r_0 r_2\cosh\psi(r_2) dr_2 + \int^1_r r_2\ln r_2 \cosh\psi(r_2) dr_2\right)dr
\end{alignedat}
\end{equation}
where in the second line we used our first change in the order of integration. We will now perform another change of the order of integration by interchanging $r_2$ with $r$. To this end notice that \{$0\leq r\leq 1$, $0\leq r_2 \leq r$\} is equivalent with \{$0\leq r_2\leq 1$ and $r_2\leq r\leq 1$\}. Also observe that \{$r\leq r_2\leq 1$, $0\leq r\leq 1$\} is equivalent with \{$0\leq r_2\leq 1$ and $0\leq r\leq r_2$\}.
We then find, after interchanging, moving $r_2\cosh\psi(r_2)$ to the front, and performing a partial integration in the first integral that
\begin{align} \label{eq:k1_final}
\begin{split}
k_1 &= -\int^1_0 r_2\cosh\psi(r_2)\left(\int^1_{r_2} r\ln r dr + \ln r_2\int^{r_2}_0 r dr\right)dr_2\\
&= -\int^1_0 r_2\cosh\psi(r_2)\left(\frac{1}{2} r^2\ln r|^1_{r_2} - \int^1_{r_2} \frac{1}{2} r dr + \frac{1}{2}r^2_2\ln r_2\right)dr_2\\
&=\frac{1}{4}\int^1_0 r_2 \left(1-r^2_2 \right)\cosh\psi(r_2) dr_2
\end{split}
\end{align}
which is the reduced form for $k_1$.

For the integral of $k_3$ our derivation follows the same scheme as for $k_1$, by changing order of integration twice. However, in this derivation also the Poisson-Boltzmann equation is involved to deal with the hyperbolic sine-function. In this case we have
\begin{equation}
\begin{alignedat}{2}
k_3 &= &&\int^1_0 r \sinh\psi(r)\left(\int^1_r\int^{r_1}_0 \frac{r_2}{r_1}\cosh\psi(r_2) dr_2 dr_1\right) dr \\
&=&&\int^1_0 r \sinh\psi(r)\left(\int^r_0\int^1_r \frac{r_2}{r_1}\cosh\psi(r_2) dr_1 dr_2 + \int^1_r\int^1_{r_2} \frac{r_2}{r_1}\cosh\psi(r_2) dr_1 dr_2\right) dr\\ 
&=&&\int^1_0 r \sinh\psi(r)\left(\int^r_0 r_2\ln r_1|^1_r \cosh\psi(r_2) dr_2 + \int^1_r r_2\ln r_1|^1_{r_2} \cosh\psi(r_2) dr_2\right)dr\\
&={}-{}&&\int^1_0 r \sinh\psi(r)\left(\ln r\int^r_0 r_2\cosh\psi(r_2) dr_2 + \int^1_r r_2\ln r_2 \cosh\psi(r_2) dr_2\right)dr\\
&={}-{}&&\int^1_0 r_2 \cosh\psi(r_2)\left(\int^1_{r_2} r\ln r\sinh\psi(r) dr + \ln r_2\int^{r_2}_0 r\sinh\psi(r) dr\right) dr_2.
\end{alignedat}
\end{equation}

Now, invoking the Poisson-Boltzmann equation, recalling that $\frac{1}{r}\frac{\partial}{\partial r}\left( r\frac{\partial\psi}{\partial r}\right) =\frac{c_v}{\lambda^2_\textrm{ref}} \sinh \psi $ by Eq. (14) from the main text 
, we then see that by partial integration we have
\begin{align} \label{eq:k3_final}
\begin{split}
k_3&=-\int^1_0 r_2 \cosh\psi(r_2)\frac{\lambda^2_\textrm{ref}}{c_v}\left(\int^1_{r_2} \ln r\textrm{ }\frac{\partial}{\partial r}\left( r\frac{\partial\psi}{\partial r}\right) dr + \ln r_2\int^{r_2}_0 \frac{\partial}{\partial r}\left( r\frac{\partial\psi}{\partial r}\right)dr\right) dr_2\\
&=-\int^1_0 r_2 \cosh\psi(r_2)\frac{\lambda^2_\textrm{ref}}{c_v}\left(r\ln r\frac{\partial\psi}{\partial r}|^1_{r_2} - \int^1_{r_2} \frac{\partial\psi}{\partial r} dr+ r_2\ln r_2\frac{\partial\psi}{\partial r_2}\right) dr_2 \\
&=-\int^1_0 r_2 \cosh\psi(r_2)\frac{\lambda^2_\textrm{ref}}{c_v}\left(-r_2\ln r_2\frac{\partial\psi}{\partial r_2} - \psi |^1_{r_2} + r_2\ln r_2\frac{\partial\psi}{\partial r_2}\right) dr_2\\
&=-\int^1_0 r_2 \cosh\psi(r_2)\frac{\lambda^2_\textrm{ref}}{c_v} \left(\psi(r_2)-\psi_\textrm{w} \right) dr_2
\end{split}
\end{align}
which is the required reduced form for $k_3$.

Now for the final result, the reduced form of $k_7$ is by far the hardest to derive. To our knowledge, the fully reduced integral for this term was not yet available. We again start by first interchanging the order of integration,
\begin{equation}
\begin{alignedat}{2}
k_7 &=&& \int^1_0 r \cosh\psi(r)\left(\int^1_r\int^{r_1}_0 \frac{r_2}{r_1}\cosh\psi(r_2) dr_2 dr_1 \right)dr \\
&=&&\int^1_0 r \cosh\psi(r)\left(\int^r_0\int^1_r \frac{r_2}{r_1}\cosh\psi(r_2) dr_1 dr_2 + \int^1_r\int^1_{r_2} \frac{r_2}{r_1}\cosh\psi(r_2) dr_1 dr_2\right) dr\\ 
&=&&\int^1_0 r \cosh\psi(r)\left(\int^r_0 r_2\ln r_1|^1_r \cosh\psi(r_2) dr_2 + \int^1_r r_2\ln r_1|^1_{r_2} \cosh\psi(r_2) dr_2\right)dr\\
&={}-{}&&\int^1_0 r \cosh\psi(r)\left(\ln r\int^r_0 r_2\cosh\psi(r_2) dr_2 + \int^1_r r_2\ln r_2 \cosh\psi(r_2) dr_2\right)dr.\\
\end{alignedat}
\end{equation}

Up until this point the steps have been equivalent to the steps for $k_1$ and $k_3$. However, in the next steps, we will only interchange the order of integration in the second term. Notice that we then obtain a symmetry in the distribution of the variables and thus the integral expressions, resulting in
\begin{align}
\begin{split}
k_7 =&-\int^1_0 r\cosh\psi(r)\ln r\int^r_0 r_2\cosh\psi(r_2) dr_2 dr - \int^1_0 r_2\cosh\psi(r_2)\ln r_2\int^{r_2}_0 r\cosh\psi(r) dr dr_2\\
=&-2 \int^1_0 r_2\cosh\psi(r_2)\ln r_2\int^{r_2}_0 r\cosh\psi(r) dr dr_2.
\end{split}
\end{align}

Now we finish our derivation by performing partial integration and invoking the Poisson-Boltzmann equation again, resulting in
\begin{align}
\begin{split}
k_7&=-2 \int^1_0 r_2\cosh\psi(r_2)\ln r_2\int^{r_2}_0 r\cosh\psi(r) dr dr_2\\
&=-2 \int^1_0 r_2\cosh\psi(r_2)\ln r_2\left(\frac{1}{2}r_2^2\cosh\psi(r_2) - \int^{r_2}_0 \frac{1}{2} r^2 \sinh\psi(r) \frac{\partial\psi}{\partial r}dr\right)dr_2 \\
&=-2 \int^1_0 r_2\cosh\psi(r_2)\ln r_2\left(\frac{1}{2}r_2^2\cosh\psi(r_2) - \frac{\lambda^2_\textrm{ref}}{c_v}\int^{r_2}_0 \frac{1}{2} r \frac{\partial}{\partial r}\left( r\frac{\partial\psi}{\partial r}\right) \frac{\partial\psi}{\partial r}dr\right)dr_2\\
&=-2 \int^1_0 r_2\cosh\psi(r_2)\ln r_2\left(\frac{1}{2}r_2^2\cosh\psi(r_2) - \frac{\lambda^2_\textrm{ref}}{2c_v}\int^{r_2}_0  \left( r \left(\frac{\partial\psi}{\partial r}\right)^2 + r^2 \frac{\partial^2\psi}{\partial r^2}\frac{\partial\psi}{\partial r} \right) dr\right) dr_2\\
&=-2 \int^1_0 r_2\cosh\psi(r_2)\ln r_2\left(\frac{1}{2}r_2^2\cosh\psi(r_2) - \frac{\lambda^2_\textrm{ref}}{2c_v}\int^{r_2}_0  \left( r \left(\frac{\partial\psi}{\partial r}\right)^2 + \frac{1}{2} r^2 \frac{\partial}{\partial r}\left(\frac{\partial\psi}{\partial r}\right)^2 \right) dr \right)dr_2.
\end{split}
\end{align}

Now we partially integrate the last term in this equation,

\begin{align} \label{eq:k7_final}
\begin{split}
k_7&=-2 \int^1_0 r_2\cosh\psi(r_2)\ln r_2\left(\frac{1}{2}r_2^2\cosh\psi(r_2) - \frac{\lambda^2_\textrm{ref}}{2c_v}\left(\int^{r_2}_0  r \left(\frac{\partial\psi}{\partial r}\right)^2 dr
+  \frac{1}{2} r^2 \left.\left(\frac{\partial\psi}{\partial r}\right)^2 \right|^{r_2}_0 -\int^{r_2}_0  r \left(\frac{\partial\psi}{\partial r}\right)^2 dr\right)\right)dr_2\\
&=-2 \int^1_0 r_2\cosh\psi(r_2)\ln r_2\left(\frac{1}{2}r_2^2\cosh\psi(r_2) - \frac{\lambda^2_\textrm{ref}}{4c_v}\left( r_2 \frac{\partial\psi}{\partial r_2}\right)^2\right)dr_2\\
\end{split}
\end{align}
which is the reduced form of the $k_7$ integral. It is very interesting to notice that reduction of this integral does not work out in the planar case (i.e., the pore consisting of two narrow plates instead of a cylinder). In that case the radial cancellations in the last five steps of $k_7$ do not work out, due to the different form of the Jacobian (being unity) and the Laplacian (containing no reciprocal terms). 

\section*{Appendix B. Full equations of motion}
Based on the original capillary pore model, Eq.~(\ref{eq:fluxesnforces}) from the main text, it is possible to obtain full $(x,r)$-dependent expressions for the three fluxes by inserting Eq.~(\ref{eq:u-expression})
into Eq.~({\ref{eq:j_ions}), and omitting the averaging step, resulting in

\begin{equation}\label{eq:fluxesforces_radial}
\left( {u_x(r)} , {j_{\mathrm{ions},x}(r)} , {j_{\mathrm{ch},x}(r)} \right)^{\textrm{t}}=\begin{pmatrix} L_{11}' & L_{12}' & L_{13}'\\  L_{21}' & L_{22}' & L_{23}'\\ L_{31}' & L_{32}' & L_{33}' \end{pmatrix}\cdot \left(  -\frac{\partial p_{t,v}}{\partial x} , -\frac{\partial \mu_v}{\partial x} ,  -\frac{\partial\phi_v}{\partial x} \right)^{\textrm{t}}
\end{equation}
where
\begin{equation}\label{eq:Lmatrix_radial}
\begin{alignedat}{2}
L_{11}' &=+ \frac{1}{4\alpha} && \left(1-r^2 \right) \\
L_{12}' &=- \frac{2c_v}{\alpha} && \left(\ln r \int^r_0  r_1 \cosh\psi(r_1) dr_1 + \int^1_r  r_1 \ln r_1\cosh\psi(r_1) dr_1\right) \\ 
L_{13}' &=+ \> \frac{2}{\alpha} && \lambda^2_\mathrm{ref}\left(\psi(r) - \psi_\mathrm{w} \right) \\
L_{21}' &=+\frac{c_v}{2\alpha} && \left(1-r^2 \right)\cosh\psi(r)\\
L_{22}' &= -\frac{4 c_v}{\alpha} && \cosh\psi(r)\left(c_v\left(\ln r \int^r_0  r_1 \cosh\psi(r_1) dr_1 
+ \int^1_r  r_1 \ln r_1\cosh\psi(r_1) dr_1\right) - \frac{\alpha}{2}\right)\\ 
L_{23}' &=+ \frac{4c_v}{\alpha} && \left(\cosh\psi(r)\lambda^2_\mathrm{ref} \left(\psi(r) - \psi_\mathrm{w} \right) - \frac{\alpha}{2}\sinh\psi(r)\right)\\
L_{31}' &=-\frac{c_v}{2\alpha} && \left( 1-r^2 \right) \sinh\psi(r)\\
L_{32}' &=+\frac{4 c_v}{\alpha} && \sinh\psi(r)\left(c_v\left(\ln r \int^r_0  r_1 \cosh\psi(r_1) dr_1 
+ \int^1_r  r_1 \ln r_1\cosh\psi(r_1) dr_1\right) - \frac{\alpha}{2}\right)\\
L_{33}' &=-\frac{4 c_v}{\alpha}&\enskip  & \left(\sinh\psi(r)\lambda^2_\mathrm{ref} \left(\psi(r) - \psi_\mathrm{w} \right) - \frac{\alpha}{2}\cosh\psi(r)\right).
\end{alignedat}
\end{equation}

Solving for these fluxes, considering the appropriate boundary conditions, yields a complete picture of the velocity fields of the ions and the solvent in the cylindrical pore.

\end{document}